\documentstyle[preprint,aps,psfig]{revtex}
\begin{document}
\tightenlines
\draft

\title{Effect of transport coefficients on the
time-dependence\\ of density matrix}
\author{Yu.V.Palchikov$^{1}$, G.G.Adamian$^{1,2}$, N.V.Antonenko$^{1,2}$  and
W.Scheid$^{2}$}
\address{$^{1}$Joint Institute for Nuclear Research, 141980 Dubna, Russia\\
$^{2}$Institut f\"ur Theoretische Physik der
Justus--Liebig--Universit\"at,
D--35392 Giessen, Germany
}
\date{\today}
\maketitle

\begin{abstract}
For Lindblad's master equation of open quantum systems
with a general quadratic form of the Hamiltonian, the
propagator of the density matrix is analytically calculated
by using path integral techniques. The time-dependent
density matrix is applied to nuclear barrier penetration
in heavy ion collisions with inverted oscillator
and double-well potentials. The quantum mechanical decoherence of
pairs of phase space histories in the propagator is studied and shown
that the decoherence depends crucially on the transport coefficients.
\end{abstract}


\section{Introduction}
In many problems of nuclear physics and quantum
optics, where one deals with open quantum systems,
the memory time of the environment is very
short and a Markovian approximation is suitable.
Disregarding the averaging over the intrinsic degrees of freedom,
one can consider the open system
starting from the general Markovian master equation
for the reduced density matrix of the collective degrees of freedom
as given by Lindblad \cite{18}
\begin{eqnarray}
\frac{d\hat\rho (t)}{dt}=-\frac{i}{\hbar}[\hat H_0,\hat\rho]+
\frac{1}{2\hbar}\sum_{j}\left([\hat V_j\hat\rho, \hat V_j^+]+
[\hat V_j, \hat\rho \hat V_j^+]\right).
\label{1_eq}
\end{eqnarray}
Here, $\hat H_0$ is the Hamiltonian of the collective subsystem and
$\hat V_j$
are operators acting in the Hilbert space of the subsystem.
The terms in the sum of Eq.~(\ref{1_eq}) are
responsible for the friction and
diffusion and supply the irreversibility
of the dynamics of the open quantum system.
Omitting these terms we get a standard form for the
evolution equation of the density matrix for closed systems.
This equation and similar equations were used e.g. in
Refs.~\cite{16,17,Dod2,21,20,22,23,25,26,14,15}.

Path integral methods are a conventional tool to describe
open quantum systems \cite{26,14,15,2,3,6,8,12,BULKU}.
Here, we use results of Strunz \cite{26} elaborated with path integral
techniques and derive analytical expressions for the time-dependent
density matrix for a general Hamiltonian of quadratic form with
an inverse oscillator potential which can be applied
to the description of fission and fusion through potential barriers in
nuclear physics.
The decoherence of pairs of phase space trajectories
will be studied in the semiclassical limit for
different choices of the effects of the environment on the system.
As was shown in \cite{14}, the initial Gaussian distribution remains to
be Gaussian in an oscillator potential. We extend this
statement for any quadratic form of the Hamiltonian of the subsystem.
By a direct numerical solution of Eq.(\ref{1_eq}) we consider
the evolution of the density matrix in time in a double-well
potential under various sets of transport coefficients.
Such potentials are more useful and realistic to investigate
nuclear fission problems than inverse oscillator potentials.

\section{Path integral propagator and decoherence}
With the propagator
$G(q,q',t;q_0,q^{'}_0,0)$ of the density matrix
one can find the density matrix $<q|\hat\rho(t)|q'>$
(in coordinate representation) at any time from the initial
one $<q|\hat\rho(t=0)|q'>$:
\begin{eqnarray}
<q|\hat\rho(t)|q'>=\int dq_0 \int dq^{'}_0
G(q,q',t;q_0,q^{'}_0,0) <q_0|\hat\rho(t=0)|q^{'}_0>.
\label{Prop_eq}
\end{eqnarray}
In the one-dimensional case an expression
for the phase space path integral of
the propagator corresponding to (\ref{1_eq})
was derived in \cite{26} as
\begin{eqnarray}
G(q,q',t;q_0,q^{'}_0,0)&=&\int\limits_{(q_0,0)}^{(q,t)}D[\alpha]
\int\limits_{(q^{'}_0,0)}^{(q',t)}D[\alpha']
\exp{(\frac{i}{\hbar}S[\alpha; \alpha'])}, \nonumber \\
S[\alpha; \alpha']&=& S[q,p; q',p']=
\int\limits_{0}^{t}d\tau \{\dot q(\tau)p(\tau) - H_{eff}(q(\tau),p(\tau))\}
\nonumber \\
&-&
\int\limits_{0}^{t}d\tau \{\dot q'(\tau)p'(\tau) -
H^{*}_{eff}(q'(\tau),p'(\tau))\}
\nonumber \\
&-&i\sum_{j}^{}\int\limits_{0}^{t}d\tau \{V_j(q(\tau),p(\tau))
 V^{*}_j(q'(\tau),p'(\tau))\},
\label{2_eq}
\end{eqnarray}
with phase space paths $[\alpha]=[q,p]$, where $q$ and $p$ are
the position and momentum, respectively.
The effective Hamiltonian is given by
$$H_{eff}=H_0 - \frac{i}{2}\sum_{j}^{}|V_j|^2.$$
Here, the quantities
$H_0$, $|V_j|^2$, $V_j$ and $V_j^*$ are
the Wigner transforms of the operators $\hat H_0$,
$\hat V^+_j\hat V_j$, $\hat V_j$ and
$\hat V^+_j$ in (1), respectively.

Choosing an inverse oscillator potential, we write the Hamiltonian
of the collective subsystem in a more general quadratic form
\begin{eqnarray}
\hat H_0=\frac{1}{2m}\hat
p^2-\frac{m\omega^2}{2}\hat q^2+\frac{\mu}{2}(\hat p \hat q+\hat q \hat p).
\label{3_eq}
\end{eqnarray}
The environment operators are assumed as linear
\begin{eqnarray}
\hat V_j=A_j\hat p +B_j\hat q,\quad
\hat V_j^+=A_j^*\hat p +B_j^*\hat q, \quad j=1,2.
\label{4_eq}
\end{eqnarray}
As shown in \cite{26} for the similar case of an harmonic oscillator,
the integrals in (\ref{2_eq}) over the
momentum yield Gaussian integrals and can be evaluated.
Then the propagator is reduced to path integrals in
coordinate space \cite{26}:
\begin{eqnarray}
G(q,q',t;q_0,q^{'}_0,0)&=&\int\limits_{q_0}^{q(t)}D[q]
\int\limits_{q^{'}_0}^{q'(t)}D[q']
\exp{(\frac{i}{\hbar}S[q;q'])},\nonumber\\
S[q;q']&=&S_{cl}[q]-S_{cl}[q']-i\hbar\lambda t+\Phi[q,q'] +\frac{i}{2}D[q,q']^2.
\label{21_eq}
\end{eqnarray}
In Eq.~(\ref{21_eq})
the classical action of the isolated system $S_{cl}$,
the phase function $\Phi[q,q']$ and the square of the decohering
amplitude $D[q,q']$ can be expressed as
\begin{eqnarray}
S_{cl}&=&\int\limits_{0}^{t} d\tau\left\{\frac{1}{2}m\dot q^2+
\frac{m}{2}\omega^2 q^2\right\},\label{22_eq}\\
\Phi[q,q']&=&m\lambda\int\limits_{0}^{t}d\tau (\dot q q'-q\dot q')+
m\frac{\lambda_p-\lambda_q}{2} \int\limits_{0}^{t}d\tau (q'\dot q'-q \dot q)-
m\frac{\lambda_p\lambda_q}{2}\int\limits_{0}^{t}d\tau(q^2-{q'}^2),
\label{Phase_eq}\\
D[q,q']^2
&=&\frac{2}{\hbar}\left\{(D_{pp}+m^2\lambda_p^2D_{qq}+
2m\lambda_pD_{pq})\int\limits_{0}^{t}d\tau(q-q')^2\right.\nonumber\\
&-&\left.2m(D_{pq}+m\lambda_pD_{qq})\int\limits_{0}^{t}d\tau(q-q')(\dot q-\dot q')
+m^2D_{qq}\int\limits_{0}^{t}d\tau(\dot q-\dot q')^2\right\}.
\label{Decoh_eq}
\end{eqnarray}
The quantum mechanical diffusion coefficients are
$D_{qq}=\frac{\hbar}{2}\sum\limits_{j}|A_j|^2$ for coordinate,
$D_{pp}=\frac{\hbar}{2}\sum\limits_{j}|B_j|^2$ for momentum and
$D_{qp}=-\frac{\hbar}{2}{\rm Re}\sum\limits_{j}A_j^*B_j$ for the mixed
case. The frictional damping rate
$\lambda=-{\rm Im}\sum\limits_{j}A_j^*B_j$ and the diffusion coefficients
must satisfy the constraint:
$D_{pp}D_{qq}-D_{pq}^2\ge \lambda^2\hbar^2/4$ with
$D_{qq}>0$, $D_{pp}>0$
which secures the non-negativity of the density matrix at any time.
The values $\lambda_p=\lambda+\mu$ and $\lambda_q=\lambda-\mu$
($\lambda_p+\lambda_q=2\lambda$) are  friction coefficients for
coordinate and momentum, respectively. Both position and momentum
undergo a direct damping and diffusion process in contrast to the classical
case. If $D[q,q']$ increases with
time for $q\ne q'$, then the propagator suppresses the non-diagonal
components of the density matrix. Thus, the
interference between different positions $q$ and $q'$ becomes weaker.

Since $H_{eff}$ depends at most quadraticly on $p$ and $q$,
the path integrals are Gaussian.
In that case a semiclassical solution of the path integrals with
the method of stationary phases leads to an exact
analytical evaluation of the propagator. First, equations of motion along
the path trajectories $q(\tau)$ and $q'(\tau)$ (complex trajectories)
are calculated with the condition of stationary phase
$\delta S[q,p;q',p']=0$ with $S$ of Eq.~(3). The following equations for
$Q_1=q+q'$, $Q_2=q-q'$, $P_1=p+p'$ and $P_2=p-p'$ result,
which are solved with the boundary conditions
${\bf q}=(q(0)=q_0,q(t),q'(0)={q_0}',q'(t))$.
\begin{eqnarray}
\left(\matrix{\dot Q_1\cr \dot P_1\cr \dot Q_2 \cr \dot P_2\cr}\right)=
\left(\matrix{-\lambda_q&m^{-1}&\frac{4iD_{pq}}{\hbar}&
\frac{-4iD_{qq}}{\hbar}\cr
m\omega^2&-\lambda_p& \frac{4iD_{pp}}{\hbar}&\frac{-4iD_{pq}}{\hbar}\cr
0&0&\lambda_p&m^{-1}\cr
0&0&m\omega^2&\lambda_q\cr}\right)
\left(\matrix{Q_1\cr P_1\cr Q_2 \cr P_2\cr}\right)
\label{Motio_eq}
\end{eqnarray}
Next, the solutions $q(\tau)$ and $q'(\tau)$ of Eq.(\ref{Motio_eq})
depending on the parameters $q_0$, $q(t)$, ${q_0}'$ and $q'(t)$
are inserted into the action function $S[q;q']$ of Eq.(6) and
integrated over $\tau$. The square of
the decohering amplitude is found as:
\begin{eqnarray}
D[q,q']^2=\sinh^{-2}[\psi t]
(A_t(q_0-q'_0)^2-A_{-t}(q(t)-q'(t))^2
+B_t(q_0-q'_0)(q(t)-q'(t))),
\label{Deccal_eq}
\end{eqnarray}
where $\psi=\sqrt{\omega^2+(\lambda_q-\lambda_p)^2/4}$ and
\begin{eqnarray}
A_t&=&\frac{D_{pp}-m(m\omega^2D_{qq}+(\lambda_q-\lambda_p)D_{pq})}
{2\hbar\lambda}\nonumber\\
&+&\exp[2\lambda t]\psi^2
\frac{D_{pp}-m(-2D_{pq}\lambda_p+mD_{qq}(\omega^2-2\lambda\lambda_p
))}{2\hbar\lambda(\lambda_p\lambda_q-\omega^2)}
\nonumber\\
&+&\frac{\psi}{2\hbar(\omega^2-\lambda_p\lambda_q)}\sinh[2\psi t]
\{D_{pp}+m(2D_{pq}\lambda_p+mD_{qq}(\omega^2-\lambda_p(\lambda_q-\lambda_p))\}
\nonumber\\
&-&\frac{1}{2\hbar(\omega^2-\lambda_p\lambda_q)}\cosh[2\psi t]
\{-\lambda D_{pp}+m[(-2D_{pq}+m(\lambda_q-\lambda_p) D_{qq})
\psi^2\nonumber\\
&-&\lambda(-D_{pq}(\lambda_q-\lambda_p)+mD_{qq}
(\omega^2+0.5(\lambda_q-\lambda_p)^2))]\},\nonumber\\
B_t&=&\frac{\psi\sinh[(\psi-\lambda)t]}
{\hbar\lambda(\omega^2-\lambda_p\lambda_q)}
\{\psi
[-D_{pp}+m(
-2D_{pq}\lambda_p+mD_{qq}(\omega^2-2\lambda\lambda_p)) ]
\nonumber\\
&-&\lambda [D_{pp}+m(
2D_{pq}\lambda_p+mD_{qq}(\omega^2-\lambda_p(\lambda_q-\lambda_p))) ]\}
\nonumber\\
&-&\frac{\psi\sinh[(\psi+\lambda)t]}
{\hbar\lambda(\omega^2-\lambda_p\lambda_q)}
\{\psi
 [-D_{pp}+m(
-2D_{pq}\lambda_p+mD_{qq}(\omega^2-2\lambda\lambda_p)) ]
\nonumber\\
&+&\lambda [D_{pp}+m(
2D_{pq}\lambda_p+mD_{qq}(\omega^2-\lambda_p(\lambda_q-\lambda_p))) ]\}.
\label{App_eq}
\end{eqnarray}
Similar analytical expressions are obtained for the action $S_{cl}$
and phase $\Phi[q,q']$. Then,
the propagator (\ref{21_eq}) is finally evaluated as
\begin{eqnarray}
G(q,q',t;q_0,q^{'}_0,0)&=&\frac{m\psi}
{2\pi \hbar \sinh(\psi t)}
\exp(\lambda t)\exp(iS_R/\hbar)
\exp(-D[q,q']^2/(2\hbar)),\nonumber\\
S_R&=&S_{cl}[q]-S_{cl}[q']+\Phi[q,q']\nonumber\\
&=&\frac{m\omega}{2\sinh(\psi t)}
\bigl\{\cosh(\psi t - \phi)
[q_0^2 - {q^{'}_0}^2]
+\cosh(\psi t + \phi)
[q^2-{q^{'}}^2] \nonumber\\
&-& 2\cosh(\phi)\cosh(\lambda t)[q_0q-q_0^{'}q']
-2\cosh(\phi)\sinh(\lambda t)[q_0q'-q_0^{'}q]\bigr\},
\label{5_eq}
\end{eqnarray}
where $\sinh\phi=(\lambda_q-\lambda_p)/(2\omega)$.
This propagator is correct for any quadratic Hamiltonian
and is a generalization of the results of Refs.~\cite{26,14,15}
where propagators were obtained for harmonic  and inverted oscillators only.
For the initial density matrix ($\bar q(0)$ and $\bar p(0)$ are mean values)
\begin{eqnarray}
<q|\hat\rho(0)|q'>&=& (2\pi \sigma_{qq}(0))^{-1/2}\nonumber\\
&\times&\exp\bigl[-\frac{1}{4\sigma_{qq}(0)}\{(q -\bar q(0))^2
+(q' -\bar q(0))^2\} -
\frac{i}{\hbar}\bar p(0)(q'-q)\bigr],
\label{71_eq}
\end{eqnarray}
the density matrix at time $t$ is calculated with (\ref{Prop_eq}) and (13)
as follows
\begin{eqnarray}
<q|\hat\rho(t)|q'>&=&\frac{1}{\sqrt{2\pi\sigma_{qq}(t)}}
\exp\biggl[-\frac{1}{2\sigma_{qq}(t)}\left(\frac{q+q'}{2}-\bar q(t)\right)^2-
\frac{1}{2\hbar^2}\left(\sigma_{pp}(t)-
\frac{\sigma_{pq}^2(t)}{\sigma_{qq}(t)}\right)(q-q')^2 \nonumber\\
&+&\frac{i\sigma_{pq}(t)}{\hbar \sigma_{qq}(t)}
\left(\frac{q+q'}{2}-\bar q(t)\right)(q-q')+
\frac{i}{\hbar}\bar p(t)(q-q')\biggr]
\label{IS_eq}
\end{eqnarray}
or in explicit form
\begin{eqnarray}
<q|\hat\rho(t)|q'>&=&
\frac{m\psi}
{2\sinh(\psi t)}\exp(\lambda t)
\frac{1}{\sqrt{2\pi \sigma_{qq}(0)}} \frac{1}{\sqrt{4f_3f_6-f^2_5}}\nonumber\\
&\times&\exp\left[-\frac{-f_2f_4f_5+f_1f_5^2+f_2^2f_6+f_3(f_4^2-4f_1f_6)}
{4f_3f_6-f_5^2}\right],
\label{RD_eq}
\end{eqnarray}
where
\begin{eqnarray}
f_1&=&\sinh^{-2}[\psi t]\bigl(
-\frac{\hbar\bar q^2(0)}{2\sigma_{qq}(0)}\sinh^2[\psi t]
+\frac{1}{2}A_{-t}(q-q')^2\nonumber\\
&+&\frac{1}{2}i m \omega \cosh[\psi t + \phi]
\sinh[\psi t](q^2-{q'}^2)\bigr)\nonumber\\
f_2&=&\sinh^{-2}[\psi t]\bigl(
\frac{2i\sigma_{qq}(0) \bar p(0)+\hbar\bar q(0)}{2\sigma_{qq}(0)}
\sinh^2[\psi t]
-\frac{1}{2}B_{t}(q-{q'})\nonumber\\
&-&i m \omega \cosh[\phi]
\sinh[\psi t]
(q\cosh[\lambda t]+{q'}\sinh[\lambda t])\bigr)\nonumber\\
f_3&=&\sinh^{-2}[\psi t]
\frac{1}{2}\bigl(
im\omega\cosh[\psi t - \phi]
\sinh[\psi t]
-\frac{\hbar}{2\sigma_{qq}(0)}\sinh^2[\psi t]
-A_{t}\bigr)\nonumber\\
f_4&=&\sinh^{-2}[\psi t]\bigl(
\frac{-2i\sigma_{qq}(0) \bar p(0)+\hbar\bar q(0)}{2\sigma_{qq}(0)}
\sinh^2[\psi t]
+\frac{1}{2}B_{t}(q-{q'})\nonumber\\
&+&i m \omega \cosh[\phi]
\sinh[\psi t]
(q\sinh[\lambda t]+{q'}\cosh[\lambda t])\bigr)\nonumber\\
f_5&=&\sinh^{-2}[\psi t] A_{t}\nonumber\\
f_6&=&-\sinh^{-2}[\psi t]\frac{1}{2}
\bigl( im\omega\cosh[\psi t -\phi]
\sinh[\psi t]
+\frac{\hbar}{2\sigma_{qq}(0)}\sinh^2[\psi t]+A_{t}\bigr).
\label{FEX_eq}
\end{eqnarray}
Here, $\bar q(t)$ and $\bar p(t)$ are the mean values
of $\hat q$ and $\hat p$, respectively, and
$\sigma_{qq}(t)$, $\sigma_{pp}(t)$ and $\sigma_{pq}(t)$ the corresponding
variances \cite{23,15}. Explicit expressions for these mean values
and variances are given in Ref.~\cite{15}.
The diagonal part of the density matrix
(\ref{IS_eq}) yields a Gaussian distribution at time $t$
\begin{eqnarray}
\rho (q,t)=<q|\hat\rho (t)|q>=
(2\pi \sigma_{qq}(t))^{-1/2}
\exp\bigl[-\frac{1}{2\sigma_{qq}(t)}(q -\bar q(t))^2\bigr],
\label{8_eq}
\end{eqnarray}
where
\begin{eqnarray}
\bar q(t)&=&
e^{-\lambda t}\left(\bar q(0)\left[\cosh(\psi t)+
\frac{\lambda_p-\lambda_q}{\psi}\sinh(\psi t)\right]+
\frac{1}{m\psi}\bar p(0)\sinh(\psi t)\right),\nonumber\\
\sigma_{qq}(t)&=&
\frac{1}{2m^2\lambda(\omega^2-\lambda_p\lambda_q)}
\left[m^2(\omega^2-2\lambda_p\lambda)D_{qq}-D_{pp}-2m\lambda_pD_{pq}\right]
\nonumber\\
&+& e^{-2\lambda t}\left[
\frac{2C_1}{m(\lambda_q-\lambda_p)}-
\frac{1}{2m\omega^2}[(\lambda_q-\lambda_p)C_2+2C_3\psi]\cosh(2\psi t)\right.
\nonumber\\
&+&\left.\frac{1}{2m\omega^2}[(\lambda_q-\lambda_p)C_3+2C_2\psi]\sinh(2\psi t)
\right]
\label{10a_eq}
\end{eqnarray}
with the following notations
\begin{eqnarray}
C_1&=&\frac{m\omega^2(\lambda_q-\lambda_p)}{4\psi^2}\biggl[
\sigma_{qq}(0)-\frac{1}{m^2\omega^2}\sigma_{pp}(0)+
\frac{\lambda_q-\lambda_p}{m\omega^2}\sigma_{pq}(0)\nonumber\\
&-&\frac{1}{\lambda}D_{qq}+\frac{1}{m^2\omega^2\lambda}D_{pp}-
\frac{(\lambda_q-\lambda_p)}{m\omega^2\lambda}D_{pq}\biggr],\nonumber\\
C_2&=&\frac{1}{4\psi^2}\biggl[
\frac{\lambda_q-\lambda_p}{m}(\sigma_{pp}(0)-
m^2\omega^2\sigma_{qq}(0))+
4\omega^2\sigma_{pq}(0)\nonumber\\
&+&\frac{1}{\omega^2-\lambda_p\lambda_q}
\biggl(\frac{2\omega^2-\lambda_p\lambda_q}
{m}[D_{pp}+
m^2\omega^2D_{qq}]+\frac{\lambda_q^2}{m}D_{pp}+
\lambda_p^2 m\omega^2D_{qq}+
4\lambda\omega^2D_{pq}\biggr)\biggr],\nonumber\\
C_3&=&-\frac{1}{2m\psi}\biggl[
m^2\omega^2\sigma_{qq}(0)+
\sigma_{pp}(0)+
\frac{1}{\omega^2-\lambda_p\lambda_q}(
\lambda_qD_{pp}+ 2m\omega^2D_{pq}+
m^2\omega^2\lambda_pD_{qq})\biggr].\nonumber
\end{eqnarray}
For the values $\lambda_p=\lambda_q=0$, $D_{pp}=D_{qq}=D_{pq}=0$,
$\sigma_{pp}(0)=\hbar^2/(4\sigma_{qq}(0))$ and
$\sigma_{qp}(0)=0$,
we obtain the same result with these expressions as in Refs.~\cite{27,Dod}.
For $\lambda_q=0$, $D_{pp}=D_{qq}=D_{pq}=0$ and
$\sigma_{pp}(0)=\hbar^2/(4\sigma_{qq}(0))$ and
$\sigma_{qp}(0)=0$, our results coincide
in the underdamped limit with the results of Ref.~\cite{4},
where tunneling was studied with
the inverted Caldirola-Kanai Hamiltonian.
For $\lambda_q=0$ and $D_{qq}=0$, our result can be transformed
to the one of Ref.~\cite{Hof}.

\section{Calculated results}
An influence of the friction and diffusion coefficients
on the tunneling was considered in Refs.~\cite{14,15,AASN}.
Here, we study the time-dependence of the decohering amplitude
and the non-diagonal part of the density matrix for different
sets of the transport coefficients.
In order to demonstrate the effect of the diffusion and friction
coefficients of the coordinate, $D_{qq}$
and $\lambda_q$, on the change of the distance
between phase space trajectories,
we take a simple expression for the diffusion coefficients:
\begin{eqnarray}
D_{pp}&=&(1+\kappa)\lambda m\hbar\omega_{eff}
\coth(\hbar\omega_{eff}/(2kT))/2, \nonumber\\
D_{qq}&=&(1-\kappa)\hbar\lambda\coth(\hbar\omega_{eff}/(2kT))/
(2m\omega_{eff}),\quad D_{pq}=0.
\label{DC_eq}
\end{eqnarray}
Here, $\kappa$ is an adjustable parameter.
The parameter $\omega_{eff}$ could be found from a
microscopic consideration of the open system.
With $\kappa=1$ and $D_{qq}=D_{pq}=0$ we obtain
the "classic" set of diffusion coefficients which does not preserve
the non-negativity of the density matrix at all times \cite{21,22,23}.

As an example we consider the relative motion of the
two nuclei $^{76}$Ge and $^{170}$Er at the
Coulomb barrier which is approximated by the inverted oscillator.
Fig.~1 shows the time-dependence of the density matrix $\rho(q,q')$
for $\kappa=0$ and 1 in (\ref{DC_eq}), $\lambda_p=(1+\kappa)\lambda$
and $\lambda_q=(1-\kappa)\lambda$.
Since with $\kappa=0$ and 1 the density matrix is practically
diagonal after a short time
interval of about $5\times 10^{-22}$ s, semiclassical methods work
quite well in heavy ion collisions.
The density matrix
becomes faster diagonal in the case $\kappa=1$ than for $\kappa=0$.
The time behaviour of the non-diagonal
components of the density matrix is evidently correlated with
the time-dependence of the decoherence $D$ which is shown in Fig.~2.
After a decrease of $D$ during a short time interval the decoherence
increases indicating a depression of the interference between
different states (trajectories).
The decoherence increases slowest for
$\kappa=0$ and more rapidly for higher temperatures.
Further, Fig.~2 shows that  the decoherence amplitude
decreases with increasing $\lambda_q$. This has the consequence that
the penetrability through the barrier increases due to a larger
interference between different states (trajectories) \cite{14,15}.

In order to show the role of $D_{qq}$ in distorting
the coherence between states, we compare
the time-dependence of $D$ in Fig.~3 for
$\kappa=0$, 0.5 and 1 in (\ref{DC_eq}) with
$\lambda_p=2\lambda$ and $\lambda_q=0$. For times
$t>5\times 10^{-22}$ s, which are of interest for physical observables,
the decoherence increases fastly for $D_{qq}=0$ ($\kappa=1$).
With $D_{qq}\ne 0$ ($\kappa<1$) the interference between
different states survives a longer time.

Eq.~(\ref{1_eq}) can also be solved by rewriting it in
a system of equations for the
matrix elements of $\hat \rho$ in some basis \cite{AASN}.
These equations can be numerically treated for arbitrary potentials.
With complete orthogonal set of basis functions $|n>$
we obtain from Eq.(\ref{1_eq}) the system of equations
for the matrix elements of density matrix $\hat \rho$:
\begin{eqnarray}
\frac{d\rho_{mn}}{dt}&=&\sum_{l}\{-\frac{i}{\hbar}
(<m|\hat H_0|l>\rho_{ln}-<l|\hat H_0|n>\rho_{ml})\nonumber\\
&+& \rho_{ml}B_{ln} +\rho_{ln} C_{ml}
+ \sum_{l'}\rho_{ll'}A_{mll'n}\},
\label{MC_eq}
\end{eqnarray}
where the coefficients are defined as follows
\begin{eqnarray}
B_{ln}&=&\sum_{l'}(D_1^-\Delta^+_{ll'}\Delta^+_{l'n}+
{D_1^+}^*\Delta^-_{ll'}\Delta^-_{l'n}-
{D_2^-}\Delta^-_{ll'}\Delta^+_{l'n}-
{D_2^+}^*\Delta^+_{ll'}\Delta^-_{l'n}),\nonumber\\
C_{ml}&=&\sum_{l'}(D_1^+\Delta^+_{l'l}\Delta^+_{ml'}+
{D_1^+}^*\Delta^-_{l'l}\Delta^-_{ml'}-
{D_2^+}\Delta^-_{l'l}\Delta^+_{ml'}-
{D_2^-}^*\Delta^+_{l'l}\Delta^-_{ml'}),\nonumber\\
A_{mll'n}&=&-(D_1^-+D_1^+)\Delta^+_{ml}\Delta^+_{l'n}-
({D_1^-}^*+{D_1^+}^*)\Delta^-_{ml}\Delta^-_{l'n}+
(D_2^-+{D_2^-}^*)\Delta^+_{ml}\Delta^-_{l'n}\nonumber\\
&+&
(D_2^++{D_2^+}^*)\Delta^-_{ml}\Delta^+_{l'n}.
\label{MCC_eq}
\end{eqnarray}
Here, $\Delta^-_{mn}=<m|a|n>$ and $\Delta^+_{mn}=<m|a^+|n>$
are the matrix elements of the creation $a^+$ and annihilation
$a$ operators, and
$D_1^\pm=0.5(D_1\pm0.5(\lambda_p-\lambda_q))$,
$D_2^\pm=0.5(D_1\pm0.5(\lambda_p+\lambda_q))$.
For the basis related to the eigenfunctions of
harmonic oscillator with the frequency $\omega$,
$D_1=(m\omega D_{qq}-D_{pp}/m\omega+2iD_{pq})/\hbar$ and
$D_2=(m\omega D_{qq}+D_{pp}/m\omega)/\hbar$.
For calculations, either the eigenfunctions
of harmonic oscillator
or the eigenfunctions  of potential $U(\hat q)$ are convenient
as complete orthogonal set of basis functions $|n>$
With the initial state of the open system determined
by the wave function $\Psi(q)$ the initial density
matrix is calculated as $\rho_{mn}(t=0)=<m|\Psi><\Psi|n>$.
With this initial condition we can solve Eq.(\ref{MC_eq})
and find the time dependence of average value
$F={\rm Tr}(\hat\rho(t) \hat F)$ of any operator $\hat F$
and of diagonal and nondiagonal elements of the density matrix.

Let us consider a system with mass  $m$=53$m_0$
($m_0$ is the mass of a nucleon) in
a symmetric double-well potential
\begin{eqnarray}
U(q)=-\frac{8\Delta U}{L^2}q^2+\frac{16\Delta U}{L^4}q^4
\label{DWS_eq}
\end{eqnarray}
with $\Delta U$=1.5 MeV, $L$=3 fm and start with an initial Gaussian
state for the density matrix
with a variance $\sigma_{qq}(0)=0.14 $ fm$^2$ in the left
well at $\bar q(0)=-1.5$ fm.
The calculated time-dependence of $\rho(q,q')$ is
presented in Fig.~4 for
$\kappa=1$ and $\kappa=0$ in (\ref{DC_eq}) with
$\lambda_p=2\lambda$ and $\lambda_q=0$.
The transition of the system to the right well
mainly occurs along the direction $q=q'$. At the same time the non-diagonal
part of the density matrix is larger with $D_{qq}\ne 0$ than
with $D_{qq}=0$.
For $D_{qq}\ne 0$, the distribution in the right well is wider  and
the transition rate between the two wells is larger \cite{AASN,291}.

\section{Summary}
Using the path integral method for
master equations of general Lindblad form
for Markovian open quantum systems,
we obtained an analytical expression for the propagator of
the density matrix of a general quadratic Hamiltonian coupled linearly
(in coordinate and momentum) with the environment.
The time-dependent diagonal and nondiagonal
elements of the density matrix in coordinate representation
were calculated as a function of different
sets of transport coefficients
for the inverted oscillator  and  double-well potentials.
At times of interest for heavy ion collisions at the Coulomb
barrier, the density matrix is practically diagonal which justifies the use
of semiclassical methods.
The time behaviour of the decoherence  crucially depends on the choice
of the friction and diffusion coefficients.
With diffusion coefficients preserving the non-negativity
of the density matrix at any time, the decoherence increases
slower than in the classical case with $D_{qq}=0$.
Therefore, the penetrability of a barrier is larger
in the case of $D_{qq}\ne 0$ due to a stronger coherence between different
states.

G.G.A. is grateful to the Alexander von Humboldt-Stiftung (Bonn) for support.
This work was supported in part by DFG and RFBR.

\newpage

\begin{figure}
\psfig{figure=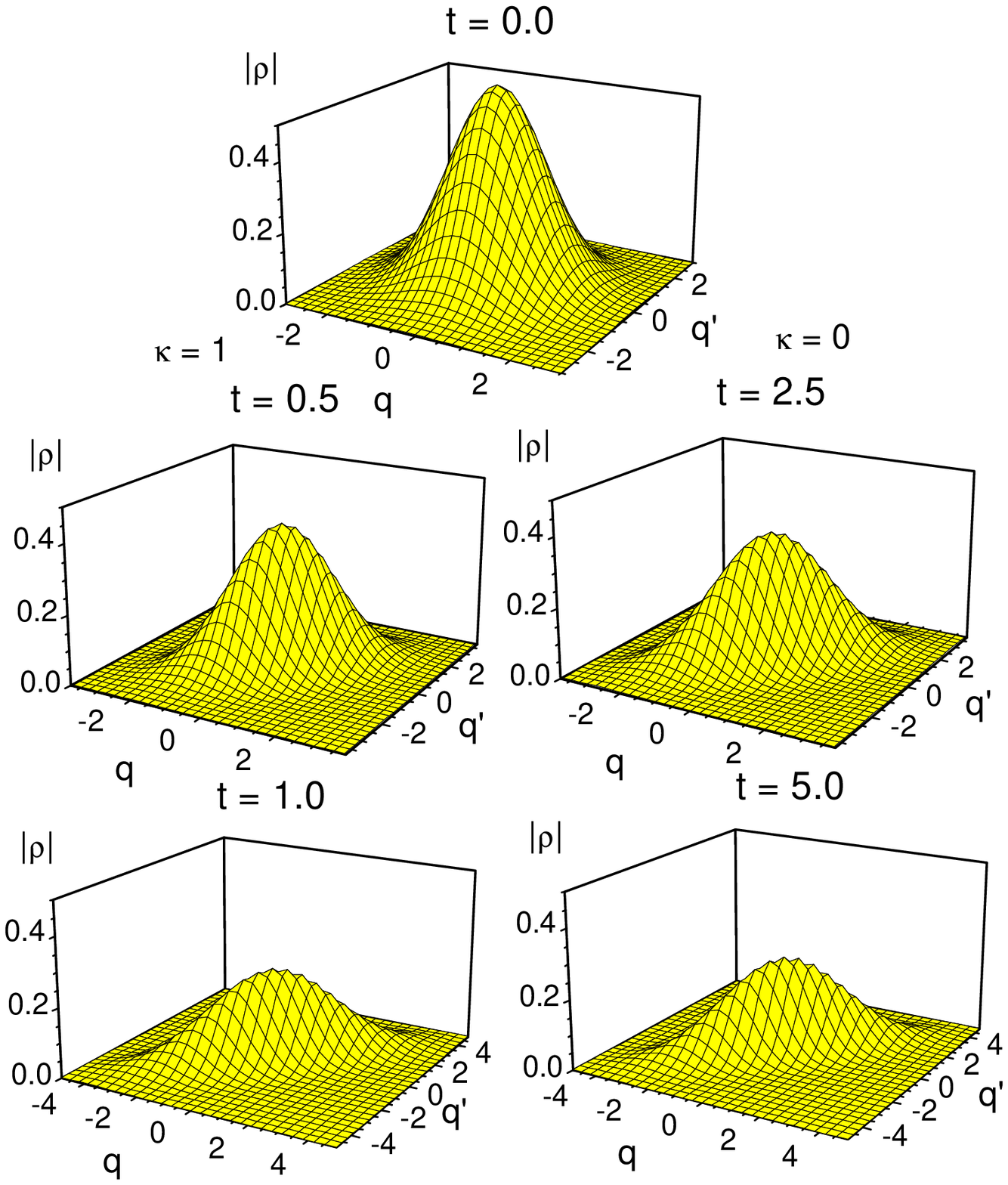,width=15cm,height=17cm}
\caption{Calculated time-dependence
of the module of the density matrix $|\rho(q,q')|$ in the
inverted oscillator potential, reproducing the Coulomb barrier
in a $^{76}$Ge+$^{170}$Er collision, for
$\kappa=1$ (left side) and 0 (right side) in (\protect\ref{DC_eq}),
$\lambda_p=(1+\kappa)\lambda$ and $\lambda_q=(1-\kappa)\lambda$.
The parameters are $\bar q(0)=0$, $\bar p(0)=0$,
$\hbar\omega=\hbar\omega_{eff}=2.0$ MeV, $\sigma_{qq}(0)=0.7$ fm$^2$,
$m=53m_0$ ($m_0$ is the mass of nucleon),
$\hbar\lambda=2$ MeV and $T=0$ MeV.
The initial density matrix is presented on the top.
The time is given in units of $6.582\times 10^{-22}$ s.}
\end{figure}
\newpage

\begin{figure}
\psfig{figure=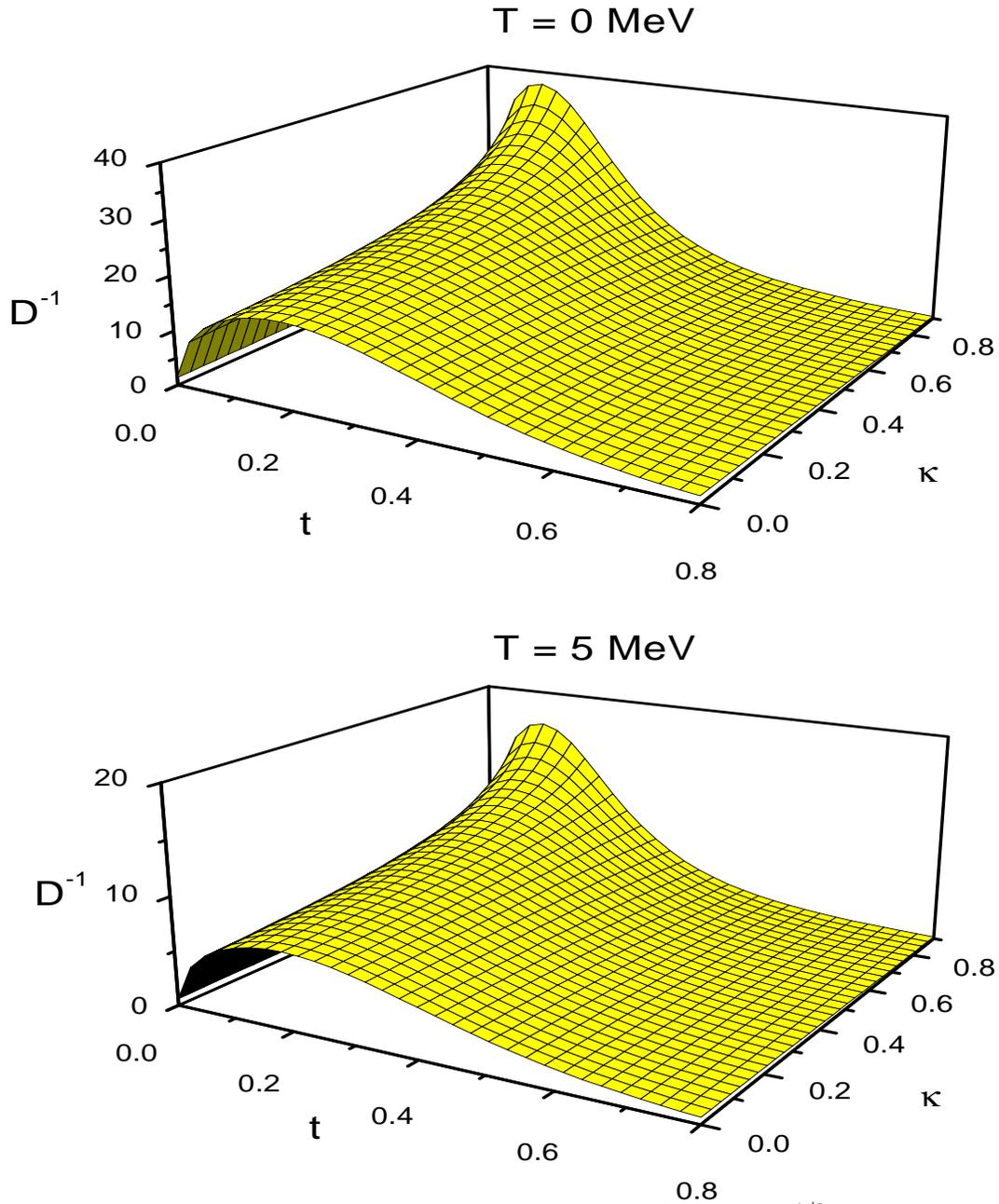,width=14cm,height=17cm}
\caption{Dependence of the inverse
decoherence $D^{-1}$ (units $\hbar^{-1/2}$) on time and $\kappa$, used
in the definitions of $D_{pp}$ and $D_{qq}$ in
(\protect\ref{DC_eq}), $\lambda_p=(1+\kappa)\lambda$ and
$\lambda_q=(1-\kappa)\lambda$ at $T=0$ and 5 MeV.
The initial values are $q(0)-q'(0)=0.01$ fm and $p(0)=p'(0)=0$.
The other parameters are the same as in Fig.1.
The time is given in units of $6.582\times 10^{-22}$ s.}
\end{figure}
\newpage

\begin{figure}
\psfig{figure=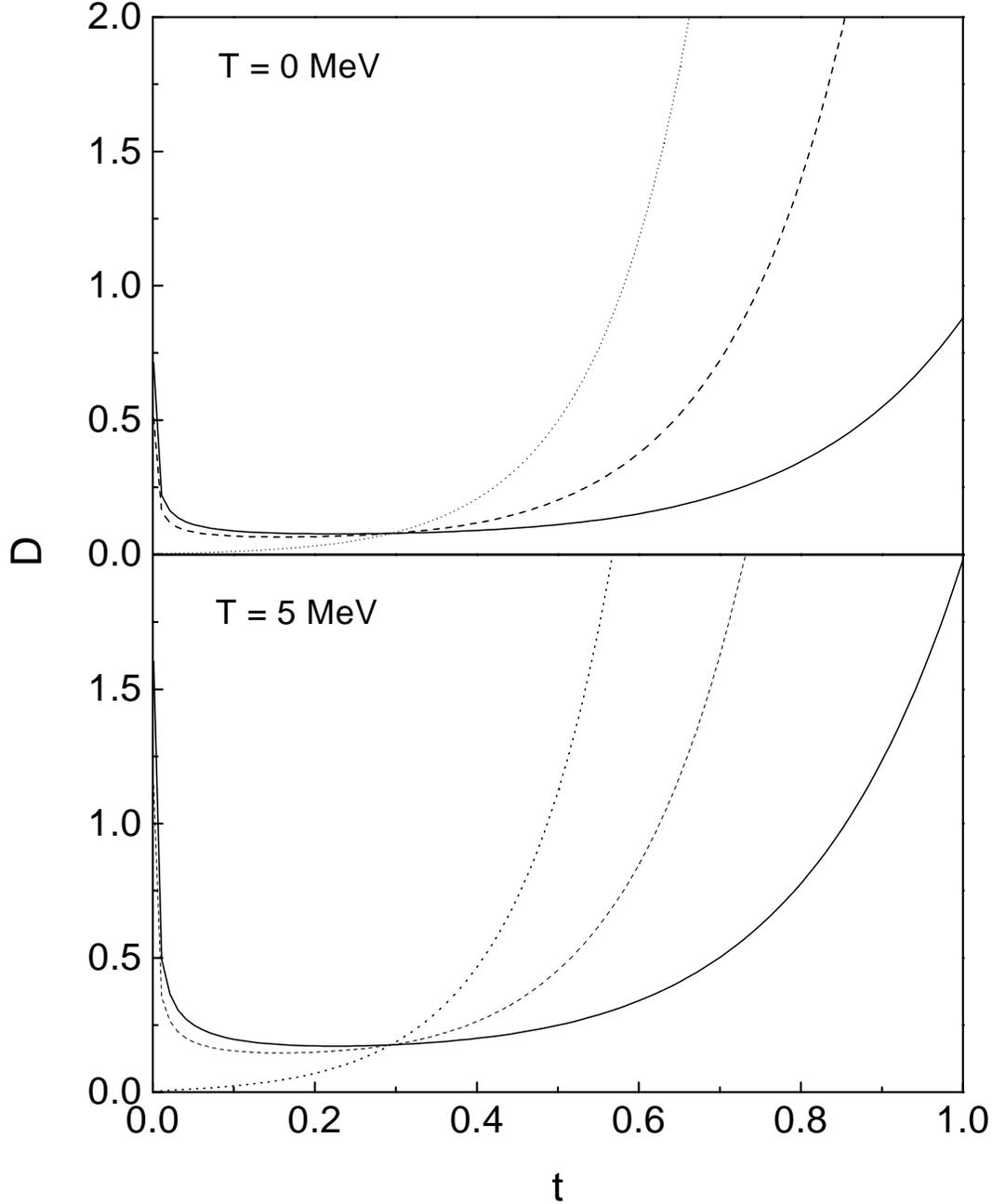,width=14cm}
\caption{Time-dependence of the
decoherence $D$ (units $\hbar^{1/2}$) for
$\lambda_p=2\lambda$, $\lambda_q=0$, $T=0$ and 5 MeV, and
$\kappa=0$ (solid lines), 0.5 (dashed lines) and 1 (dotted lines).
The initial values are $q(0)-q'(0)=0.01$ fm and $p(0)=p'(0)=0$.
The other parameters are the same as in Fig.1.
The time is given in units of $6.582\times 10^{-22}$ s.}
\end{figure}
\newpage

\begin{figure}
\psfig{figure=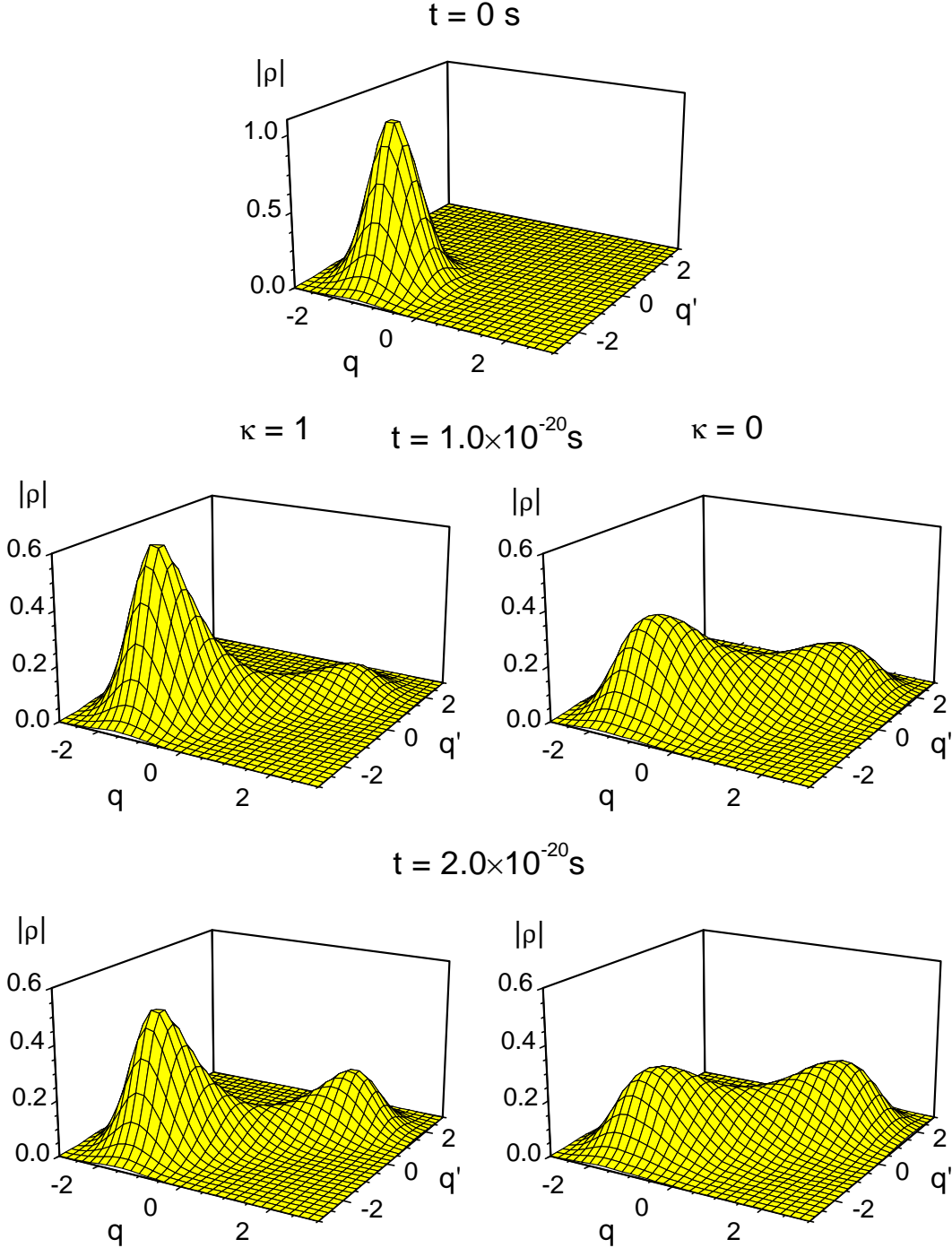,width=14cm}
\caption{Calculated time-dependence
of the module of the density matrix $|\rho(q,q')|$ in a double-well
potential (see text) for $\kappa=1$ (left side) and 0
(right side) in (\protect\ref{DC_eq}) at $T=0$ MeV
and $\lambda_p=2\lambda$, $\lambda_q=0$.
The initial Gaussian distribution (top part) with
$\sigma_{qq}(0)=0.14$ fm$^2$, $\bar p(0)=0$ starts with
$\bar q=-1.5$ fm in the left well.
The further parameters are
$\hbar\omega_{eff}=2.0$ MeV, $\hbar\lambda=2$ MeV and $m=53m_0$.
The time is indicated.}
\end{figure}

\end{document}